\documentclass{iopart}

\usepackage{ulem}
\usepackage{euscript}
\usepackage{epsfig}
\newcommand{\boldsymbol}[1]{\mbox{\boldmath $#1$}}

\begin{document}
% Journal identifier can be put here if required, e.g.
%\jl{14}

\title[ ]{Phase diagrams for an ideal gas mixture of fermionic atoms and bosonic molecules}
\author{J. E. Williams, N. Nygaard and C. W. Clark}

\address{Electron and Optical Physics Division, National Institute of Standards and Technology, Gaithersburg, Maryland 20899-8410}

\begin{abstract}
We calculate the phase diagrams for a harmonically trapped ideal gas mixture of fermionic atoms and bosonic molecules in chemical and thermal equilibrium, where the internal energy of the molecules can be adjusted relative to that of the atoms by use of a tunable Feshbach resonance. We plot the molecule fraction and the fraction of Bose-condensed molecules as functions of the temperature and internal molecular energy. We show the paths traversed in the phase diagrams when the molecular energy is varied either suddenly or adiabatically. Our model calculation helps to interpret the {\it{adiabatic}} phase diagrams obtained in recent experiments on the BEC-BCS crossover, in which the condensate fraction is plotted as a function of the {\it{initial}} temperature of the Fermi gas measured before a sweep of the magnetic field through the resonance region.
\end{abstract}

\maketitle

\section{Introduction}
\begin{figure}
\begin{center} \includegraphics[scale=0.27]{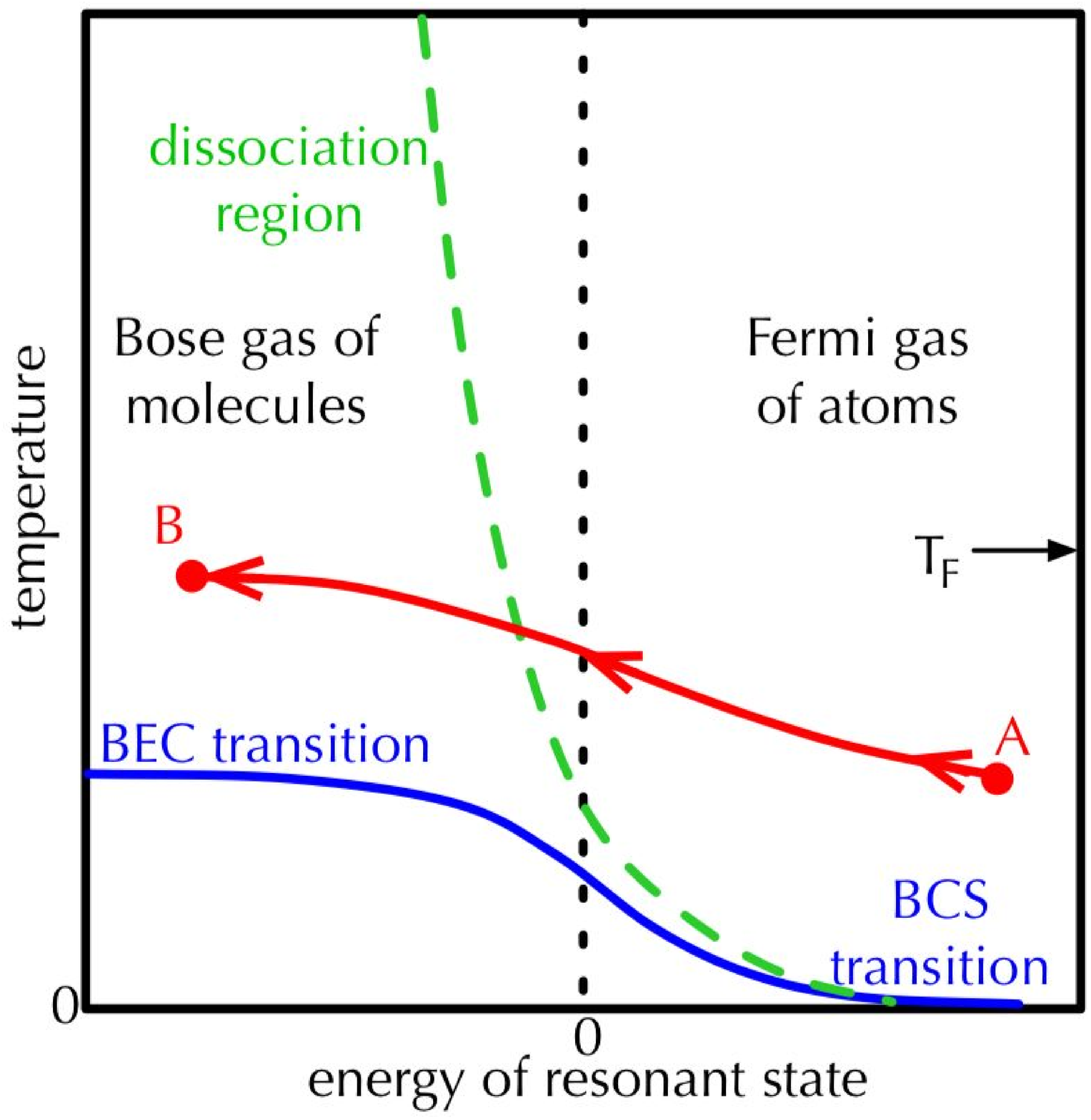}
\caption{Illustration of the atom-molecule phase diagram for a Fermi gas with resonant interactions. Starting at some initial point in the phase diagram such as $A$, the system can be continuously transformed from a gas of fermionic atoms to a gas of bosonic molecules at some point $B$ by adjusting the resonant state energy.}
\label{Fig_phasediag}
\end{center}
\end{figure}

In recent experiments, a Feshbach resonance is utilized to explore the crossover from Bardeen-Cooper-Schrieffer (BCS) superfluidity in a dilute Fermi gas to Bose-Einstein condensation (BEC) in a dilute gas of diatomic molecules \cite{Regal04a,Bartenstein03a,Zwierlein04a,Bourdel04a}. A qualitative sketch of the phase diagram for this system is shown in Figure~\ref{Fig_phasediag}. By tuning the resonant state energy, the system can be continuously transformed from a gas of fermionic atoms, starting for example at point $A$ and moving along some path to a gas of bosonic diatomic molecules at point $B$. The specific path taken depends on the rate at which the sweep is performed. 

In this article, we study the two limiting cases of an adiabatic sweep or a sudden change in order to gain a qualitative understanding of recent experiments. These two limits correspond respectively to paths of constant entropy or constant energy in the phase diagram; the final molecule fraction and condensate fraction depend strongly on which of these paths is taken. The interpretation of the recent experiments of Regal {\it{et al.}}~\cite{Regal04a} and Zwierlein {\it{et al.}}~\cite{Zwierlein04a} requires specific consideration of the time dependence of the sweep. This is because the only temperature measurements reported are those taken before the onset of the sweep. These experiments appear to have been done in the adiabatic regime. Consequently it is crucial to plot theoretical predictions versus the initial temperature or entropy in an {\it{adiabatic}} phase diagram, rather than the more traditional one (e.g. Figure~\ref{Fig_phasediag}) used in previous interpretations of these experiments~\cite{Falco04a,Diener04a}. Our simple model for a trapped system yields an adiabatic phase diagram in qualitative agreement with experimental data in the BEC limit, where a critical {\it{initial}} temperature of $T_{i,c} \approx 0.2 T_{\rm{F}}$ is found. Here $T_{\rm{F}}$ is the Fermi temperature. 

Before presenting our model and results, it is useful to place our work in the context of previous work done on this system. There are two main aspects of the problem we address that have been treated essentially as separate questions in the literature. The first question has to do with the formation of molecules as the resonance energy is varied in time \cite{Mies02a,Julienne04,Drummond1998a,Timmermans1999a,Kokkelmans2002a,Mackie2002a,Kohler2003b,Javanainen2004a,Andreev2004a} and the second issue is the determination of the equilibrium phase diagram \cite{Ohashi02a,Milstein2002a,Falco04a,Diener04a,Stajic2004a,Ohashi2003a,Perali04a}. 

\begin{figure}
\begin{center} \includegraphics[scale=0.3]{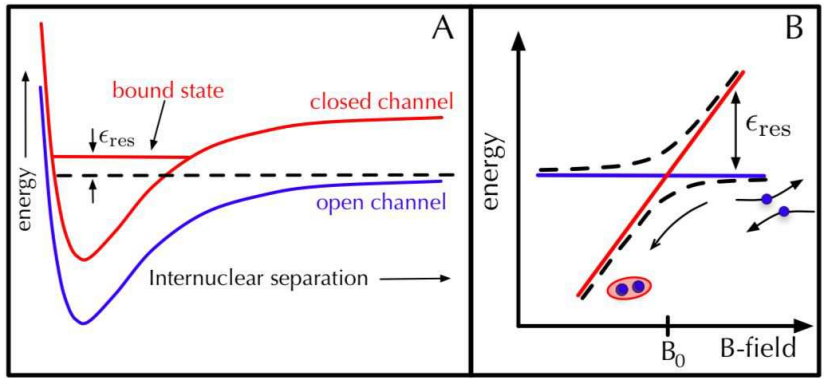}
\caption{Illustration of a Feshbach resonance. Panel A shows that a bound state in a closed  channel can be in resonance with the scattering threshold of the open channel. Panel B illustrates a two-level model to describe the quantum dynamics of a pair of colliding atoms, where one level corresponds to a scattering state and the second level is the bound state of the closed channel.}
\label{Fig_feshbach}
\end{center}
\end{figure}

A Feshbach resonance occurs between two atoms when the energy $\epsilon_{\rm{res}}$ of a closed-channel bound state coincides with the relative energy of the atoms \cite{Stwalley76,Tiesinga93a,Moerkijk95a}.  A diagram of a Feshbach resonance is shown in Figure~\ref{Fig_feshbach}A. The difference in magnetic moments between the open and closed channels allows the resonance to be tuned by adjusting an external magnetic field $B$. Due to the exchange interaction between valence electrons of the colliding pair of atoms, a coupling exists between the two spin channels. Due to this coupling, the pair of atoms spends a finite time in this closed channel bound state during the collision. This has a dramatic effect on the atomic interactions in the gas and provides a way for pairs of atoms to form a bound state, even in the absence of a third atom influencing the collision.

Most of the work on the formation of Feshbach molecules is formulated to address highly nonequilibrium quantum dynamics that occur on a timescale much shorter than the time required for the system to relax to thermodynamic equilibrium. The starting point in such an approach is to first understand the process of molecule formation of an isolated pair of colliding atoms~\cite{Mies02a,Julienne04}. One can treat the system as a two-level system, where one level corresponds to a superposition of scattering states in the open channel and the second level is the bound state of the closed channel, as illustrated in Figure~\ref{Fig_feshbach}B. The resulting coupled equations of motion can then be used to obtain a Landau-Zener adiabaticity criterion for establishing how slowly the $B$-field must be ramped to achieve a full transfer to the bound state~\cite{Mies02a,Julienne04}. More advanced treatments including many body effects have so far focussed on a gas of bosonic atoms near zero temperature~\cite{Drummond1998a,Timmermans1999a,Kokkelmans2002a,Mackie2002a,Kohler2003b}. In these theories, the system is typically far from thermodynamic equilibrium. This treatment is justified, since most of the experiments on Feshbach resonances in Bose gases are forced to probe the system for a very short time due to inelastic collisions, which heat up the gas sample~\cite{Donley02a,Herbig03a,Xu2003a,Durr2004a,Mukaiyama2004a}.

In a Fermi gas, however, the same type of inelastic processes are suppressed by several orders of magnitude \cite{Esry01,Petrov03a,Petrov03b}, resulting in a very long lifetime of the gas sample \cite{Regal03a,Strecker03a,Cubizolles03a,Jochim03b}. This permits experiments to be carried out over a time long enough for the system to relax to thermal and chemical equilibrium. In this case, a statistical description may be more appropriate than a microscopic description. The dynamics of molecule formation can then be treated using a kinetic theory approach with Boltzmann-type equations that describe the collisional relaxation processes, which set the timescale for adiabaticity~\cite{Williams04a}. 

In the thermodynamic approach, a good starting point is to first map out the equilibrium phase diagram of a Fermi gas with resonant interactions, a qualitative sketch of which is shown in Figure 1. The first calculations of the phase diagram for this type of system were done by Eagles~\cite{Eagles69a}, Leggett~\cite{Leggett80a}, and Nozi\`eres and Schmitt-Rink~\cite{Nozieres85a} using a single channel model for a homogeneous Fermi gas. This work was improved upon later by Hausmann \cite{Haussmann93a} and Chen {\it{et al.}} \cite{Chen1998a}. More recently, the phase diagram was calculated using a two-channel model for both homogeneous~\cite{Ohashi02a,Milstein2002a,Falco04a,Diener04a,Stajic2004a} and trapped gases~\cite{Ohashi2003a,Perali04a}. One of the main goals of all of these calculations is to determine the transition temperature separating the superfluid and normal phases, denoted by the blue line in Figure 1.  The important question of how the system traverses from one point in the phase diagram to another as the resonance energy is varied was not addressed in these calculations. 

To address the question of molecule formation within equilibrium thermodynamics, one can consider the two limiting cases of a very slow sweep of $\epsilon_{\rm{res}}$ or a sudden jump. If $\epsilon_{\rm{res}}$ is varied at a rate much lower than the collisional relaxation rate, the system will remain in equilibrium and the process will be adiabatic~\cite{Williams04a}. In this case the entropy is conserved during the sweep. Some aspects of this case were explored by Kokkelmans et al. for a normal gas~\cite{Kokkelmans04a} and by Carr {\it{et al.}} in the superfluid regime~\cite{Carr04a,Carr2004b}. In the other limit of changing $\epsilon_{\rm{res}}$ suddenly, on a timescale much shorter than the relaxation time, the entropy increases. In this case, the energy is conserved as the system relaxes to a new temperature. Chin and Grimm discussed this case at length, treating an ideal gas mixture of atoms and molecules in the classical regime $T\gg T_c$~\cite{Chin03b}. 

In this paper we present a detailed study of both cases of adiabatic and sudden variations in $\epsilon_{\rm{res}}$. Our work extends the work of Chin and Grimm into the quantum degenerate regime. Our simple model, which neglects interactions, describes BEC of molecules but does not allow for BCS pairing superfluidity on the $\epsilon_{\rm{res}}>0$ side of the resonance. The primary goal of our work is to develop a conceptual understanding of molecule formation in the adiabatic and sudden limits within the framework of statistical mechanics. For this purpose the ideal gas is an excellent toy model.  

Our paper is outlined as follows: in Section II we present the equilibrium theory for an ideal gas mixture of fermionic atoms and bosonic molecules that are in thermal and chemical equilibrium. Using this formulation, we calculate the phase diagrams in Section III, showing the molecule fraction and the fraction of condensed molecules versus the temperature $T$ and the resonant energy $\epsilon_{\rm{res}}$. In IIIA we consider the adiabatic limit, and in IIIB the sudden case. We then close with some concluding remarks in section IV.

\section{Equilibrium theory}
We consider an ideal gas mixture of (pseudo) spin-$1/2$ fermionic atoms and zero spin bosonic molecules in an anisotropic harmonic trap with frequencies $\omega_x$, $\omega_y$, and $\omega_z$ along the respective axes. We assume the trapping force on the molecules is twice that of the force on the atoms, so that the atomic and molecular frequencies are identical~\cite{Zwierlein03a}. The equilibrium distributions for the atoms and molecules are given by Fermi-Dirac and Bose-Einstein distributions, respectively
\begin{eqnarray}
\label{fA}
f_\sigma(\epsilon) &=& \frac{1}{e^{(\epsilon - \mu_\sigma)/k_{\rm{B}}T_\sigma} + 1} , \\
\label{fM}
f_m(\epsilon) &=& \frac{1}{e^{(\epsilon + \epsilon_{\rm{res}} - \mu_m)/k_{\rm{B}}T_m} - 1}.
\end{eqnarray}
where $\sigma \in \{\uparrow,\downarrow\}$ signifies the spin state of the atoms and $k_{\rm{B}}$ is Boltzmann's constant. In this work we restrict our attention to an equal spin mixture $\mu_\uparrow=\mu_\downarrow \equiv \mu_a$ and $T_\uparrow=T_\downarrow \equiv T_a$, so that $f_\uparrow=f_\downarrow \equiv f_a$. We assume the atoms and molecules are in thermal equilibrium so that the temperatures are equal $T_a=T_m\equiv T$. We also assume that the system has relaxed to chemical equilibrium due to collisions that convert pairs of atoms into molecules and vice versa~\cite{Williams04a,Chin03b}. In this case $\mu_m = 2\mu_a \equiv 2\mu$. The energy of a molecule is displaced relative to that of the atoms by $\epsilon_{\rm{res}}$, the magnitude and sign of which can be adjusted.

To obtain the phase diagrams, we need to calculate the population, energy, and entropy for each component. The atom population $N_a(T,\mu)$ (in a single spin component) and {\it{noncondensed}} molecule population $\tilde N_m(T,\mu,\epsilon_{\rm{res}})$ are 
\begin{eqnarray}
N_a(T,\mu) &=& \int_0^\infty d\epsilon \rho(\epsilon) f_a(\epsilon), \\
\tilde N_m(T,\mu,\epsilon_{\rm{res}}) &=& \int_0^\infty d\epsilon \rho(\epsilon) f_m(\epsilon) .
\end{eqnarray}
The density of states in a harmonic trap is
\begin{equation}
\rho(\epsilon) = \frac{1}{2} \frac{\epsilon^2}{(\hbar \bar \omega)^3} ,
\end{equation}
where $\bar \omega \equiv (\omega_x\omega_y\omega_z)^{1/3}$. Our semi-classical formulation, which replaces the sum over discrete states by an integral, is strictly valid when $k_{\rm{B}}T\gg\hbar\bar\omega$. When this inequality is not satisfied, shell effects in the Fermi sea can play a role~\cite{Schneider98a,Bruun98a}. The integrals can be evaluated and expressed in terms of Fermi $\mathcal F_n(z)$ and Bose $\mathcal G_n(z)$  integrals defined by 
\begin{eqnarray}
\mathcal F_n(z) &\equiv& \frac{1}{\Gamma(n)} \int_0^\infty \frac{x^{n-1}dx}{z^{-1} e^x + 1}, \\
\mathcal G_n(z) &\equiv& \frac{1}{\Gamma(n)} \int_0^\infty \frac{x^{n-1}dx}{z^{-1} e^x - 1} ,
\end{eqnarray}
where $\Gamma(n)$ is the gamma function. The Bose function is equal to the Riemann-zeta
function  $\zeta(n)$ when $z=1$, i.e. $\mathcal G_n(z=1) = \zeta(n)$.

In terms of these functions, the populations are
\begin{eqnarray}
\label{Na}
N_a(T,\mu) &=& \Big ( \frac{k_{\rm{B}} T}{\hbar \bar \omega} \Big )^3 \mathcal F_3(z_a), \\
\tilde N_m(T,\mu,\epsilon_{\rm{res}}) &=& \Big ( \frac{k_{\rm{B}} T}{\hbar \bar \omega} \Big ) ^3\mathcal G_3(z_m),
\label{Nm}
\end{eqnarray}
where the atom and molecule fugacities are defined by
\begin{eqnarray}
z_a(T,\mu) &\equiv& e^{\mu/k_{\rm{B}}T},\\
z_m(T,\mu,\epsilon_{\rm{res}}) &\equiv& e^{(2\mu-\epsilon_{\rm{res}})/k_{\rm{B}}T}.
\end{eqnarray}
The atom energy $E_a(T,\mu)$ and noncondensed molecule energy $\tilde E_m(T,\mu,\epsilon_{\rm{res}})$ are
\begin{eqnarray}
E_a(T,\mu) &=& \int_0^\infty d\epsilon  \rho(\epsilon) \, \epsilon \, f_a(\epsilon) \nonumber \\
&=& 3N_a   k_{\rm{B}} T \, \frac{\mathcal F_4(z_a)}{\mathcal F_3(z_a)} 
\\
\label{Ea}
\tilde E_m(T,\mu,\epsilon_{\rm{res}}) &=& \int_0^\infty d\epsilon  \rho(\epsilon) (\epsilon + \epsilon_{\rm{res}})  f_m(\epsilon) \nonumber \\
&=& 3\tilde N_m   k_{\rm{B}} T \, \frac{\mathcal G_4(z_m)}{\mathcal G_3(z_m)} + \epsilon_{\rm{res}} \tilde N_m
\label{Emn}
\end{eqnarray}
Finally, the atom entropy $S_a(T,\mu)$ and noncondensed molecule entropy $\tilde S_m(T,\mu,\epsilon_{\rm{res}})$ are
\begin{eqnarray}
\label{Sa}
S_a &=& - k_{\rm{B}} \int_0^\infty d \epsilon \rho(\epsilon) \big [ f_a \ln f_a +
(1 - f_a) \ln (1 - f_a) \big ] \nonumber \\
&=& k_{\rm{B}}N_a \Big [ 4 \frac{\mathcal F_4(z_a)}{\mathcal F_3(z_a)} - \frac{\mu}{k_{\rm{B}}T}
\Big ],\\
\tilde S_m &=& - k_{\rm{B}} \int_0^\infty d \epsilon \rho(\epsilon) \big [ f_m \ln f_m - (1 + f_m) \ln (1 + f_m) \big ] \nonumber \\
&=& k_{\rm{B}}\tilde N_m \Big [ 4 \frac{\mathcal G_4(z_m)}{\mathcal G_3(z_m)} - \frac{(2\mu-\epsilon_{\rm{res}})}{k_{\rm{B}}T}
\Big ].
\label{Sm}
\end{eqnarray}

At a low enough temperature, the molecules form a Bose-Einstein condensate. Above the transition temperature $T>T_c(\epsilon_{\rm{res}})$, $\mu < \epsilon_{\rm{res}}/2$ and the population of condensed molecules is zero. In this case the total population of atoms is
\begin{equation}
\label{Ntot}
N_{\rm{tot}}(T,\mu,\epsilon_{\rm{res}}) = 2N_a(T,\mu) + 2\tilde N_m(T,\mu,\epsilon_{\rm{res}}),
\end{equation}
where the factor of $2$ in the first term describes the two spin components and in the second term there are two atoms in each molecule. For a given value of the population $N$, the chemical potential is obtained at a specified temperature $T$ and resonance energy $\epsilon_{\rm{res}}$ by solving for $\mu$ in the equation
\begin{equation}
N_{\rm{tot}}(T,\mu,\epsilon_{\rm{res}})=N.
\end{equation}
Below $T_c(\epsilon_{\rm{res}})$, $\mu = \epsilon_{\rm{res}}/2$, so that $z_m=1$. The condensate population $N_{mc}(T,\epsilon_{\rm{res}})$ is
\begin{equation}
\label{Nmc}
N_{mc}(T,\epsilon_{\rm{res}})= N/2 - N_{\rm{tot}}(T,\mu=\epsilon_{\rm{res}}/2,\epsilon_{\rm{res}})/2.
\end{equation}
Below $T_c$ the noncondensed fraction $\tilde N_m$ reduces to the expression
\begin{equation}
\label{NmbelowTc}
\tilde N_m(T,\epsilon_{\rm{res}}/2,\epsilon_{\rm{res}}) =
\zeta(3) \Big ( \frac{k_{\rm{B}}T}{\hbar \bar \omega} \Big )^3.
\end{equation}  
The energy of condensed molecules is $\epsilon_{\rm{res}}N_{mc}(T,\epsilon_{\rm{res}})$ and the condensate entropy is zero.

On the positive detuning side of the resonance $\epsilon_{\rm{res}}>0$, we can obtain an analytic expression for $T_c$ by approximating the population of atoms by the $T=0$ limit~\cite{Pethick2002a}
\begin{equation}
\label{NaT0}
N_a(T=0,\mu=\epsilon_{\rm{res}}/2) = \frac{1}{6} \Big ( \frac{\epsilon_{\rm{res}}}{2 \hbar \bar \omega} \Big )^3.
\end{equation}
Substituting Eq.~(\ref{NmbelowTc}) and Eq.~(\ref{NaT0}) into Eq.~(\ref{Nmc}), setting 
$N_{mc}=0$, and solving for $T=T_c$, we obtain
\begin{equation}
\label{Tcpos}
T_c = \frac{\hbar \bar \omega}{\zeta(3)^{1/3}k_{\rm{B}}}
\Big [ \frac{N_{\rm{tot}}}{2} - \frac{1}{6} \Big ( \frac{\epsilon_{\rm{res}}}{2\hbar\bar\omega} \Big )^3
\Big ]^{1/3}.
\end{equation}
Taking $T_c=0$ in Eq.~(\ref{Tcpos}), we can solve for the critical value $\epsilon_{c}$, below which condensation occurs at zero temperature
\begin{equation}
\label{eresc}
\epsilon_{c} = 2 k_{\rm{B}} T_{\rm{F}}.
\end{equation}
The Fermi temperature $T_{\rm{F}}$ in a trap is defined by $k_{\rm{B}}T_{\rm{F}}=(3N)^{1/3}\hbar \bar \omega$. On the negative detuning side $\epsilon_{\rm{res}}<0$, in the limit $|\epsilon_{\rm{res}}|/k_{\rm{B}}T \gg 1$, we may set $N_a=0$ in Eq.~(\ref{Nmc}). Solving
for $T_c$ we obtain the expected result~\cite{Ohashi2003a}
\begin{equation}
\label{Tcneg}
\frac{T_c}{T_{\rm{F}}} = [6 \, \zeta(3)]^{-1/3} \approx 0.518.
\end{equation}
Equations (\ref{eresc}) and (\ref{Tcneg}) define the right and top boundaries of the condensation region in the phase diagram for an ideal gas mixture of fermionic atoms and bosonic molecules.

\section{Phase diagrams}

We now turn to the numerical calculation of the molecule fraction $\eta_m(\epsilon_{\rm{res}},T)$ and fraction of condensed molecules $\eta_{mc}(\epsilon_{\rm{res}},T)$ defined by
\begin{eqnarray}
\eta_m &=& \frac{2N_m}{N} ,\\
\eta_{mc} &=& \frac{N_{mc}}{N_m},
\end{eqnarray}
where $N_m = \tilde N_m+N_{mc}$ is the total molecule population. In our calculations, we fix the total population $N$. Then, for a given value of $T$ and $\epsilon_{\rm{res}}$, we numerically solve for the value of $\mu$ that satisfies Eq.~(\ref{Ntot}). As we approach the condensation region, the chemical potential $\mu \rightarrow \epsilon_{\rm{res}}/2$. Once we cross into the condensation region, the condensate population is calculated according to Eq.~(\ref{Nmc}).
\begin{figure}  %[htbp]
\begin{center} \includegraphics[scale=0.29]{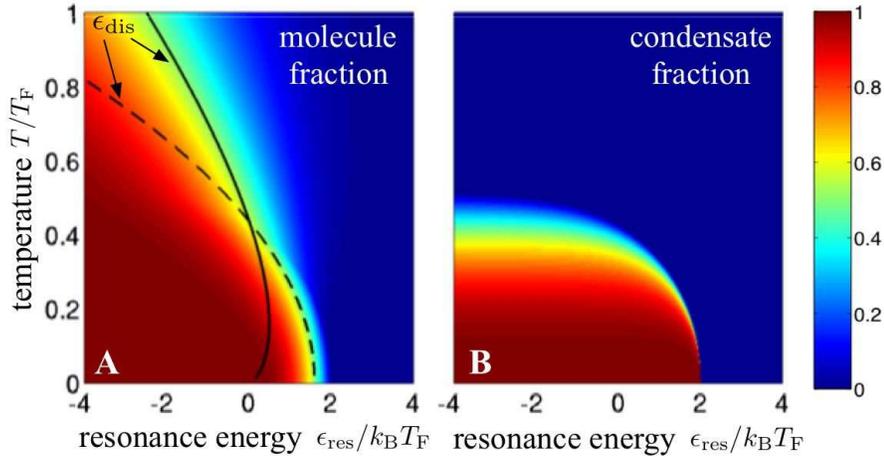}
\caption{Phase diagrams. The molecule
fraction $\eta_m(\epsilon_{\rm{res}},T)$ is plotted in panel A and the condensate fraction
$\eta_{mc}(\epsilon_{\rm{res}},T)$ is shown in panel B. The color scale is shown on the right. In panel A, the high (solid line) and low (dashed line) temperature limits of $\epsilon_{\rm{dis}}(T)$ are shown.}
\label{Fig_phasediags}
\end{center}
\end{figure}

The results of the calculations are shown in Figure~\ref{Fig_phasediags}. We plot the molecule fraction $\eta_m(\epsilon_{\rm{res}},T)$ in Figure~\ref{Fig_phasediags}A versus $T$ and $\epsilon_{\rm{res}}$, where dark blue corresponds to no molecules in the gas and dark red means there are only molecules. We also plot the condensate fraction $\eta_{mc}(\epsilon_{\rm{res}},T)$ in Figure~\ref{Fig_phasediags}B. The energy scale is set by the Fermi energy $k_{\rm{B}}T_{\rm{F}}$. By normalizing both axes by this energy, the resulting phase diagrams for an ideal gas mixture of harmonically trapped atoms and molecules are "universal", in the sense that they look identical for any set of physical parameters used in the calculation (i.e. population $N$, mass $m$, and trap frequency $\bar\omega$) . Our calculation of the condensate fraction agrees with the prediction that the condensation region is bounded on the right by $\epsilon_{c} = 2 k_{\rm{B}} T_{\rm{F}}$ and on the top by $T_c  \approx 0.518 T_{\rm{F}}$. 

In Figure~\ref{Fig_phasediags}A, the green region signifies a dissociation region where half of the atoms have been converted to molecules. We define $\epsilon_{\rm{dis}}(T)$ as the value of $\epsilon_{\rm{res}}$ where $\eta_m=0.5$. 
In recent experiments, this dissociation region is associated with the position of the resonance~\cite{Regal03a,Strecker03a}. Our calculations for an ideal gas demonstrate that $\epsilon_{\rm{dis}}(T)$ depends on the temperature. We can obtain an analytic expression for  $\epsilon_{\rm{dis}}(T)$ in the two limits of low or high temperature. In the low temperature limit, we use the Sommerfeld expansion of $\mathcal{F}_n(e^x)$ for $x\gg1$~\cite{Pathria1972a}
\begin{equation}
\mathcal{F}_n(e^x)=\frac{x^n}{\Gamma(n+1)}\Big [1 + n(n-1)\frac{\pi^2}{6}x^{-2} + \cdots \Big ].
\label{Sommer}
\end{equation}
Using Eq.~(\ref{Sommer}), we can solve for $\epsilon_{\rm{dis}}$ in the equation $N_a(T,\epsilon_{\rm{dis}}/2)=N/4$
\begin{equation}
\epsilon_{\rm{dis}}(T\ll T_{\rm{F}})/k_{\rm{B}}=2^{2/3}T_{\rm{F}} - \frac{2^{4/3}\pi^2}{3}\frac{T^2}{T_{\rm{F}}}.
\end{equation}
In the limit of $T \gg T_{\rm{F}}$, the atom and molecule distributions go over to the Maxwell-Boltzmann distributions and it is straightforward to verify that
\begin{equation}
\epsilon_{\rm{dis}}(T\gg T_{\rm{F}})/k_{\rm{B}} = -\big[3 \ln (T/T_{\rm{F}}) + \ln 12 \big] T.
\end{equation}
The high and low temperature limits of $\epsilon_{\rm{dis}}(T)$ are shown in Figure~\ref{Fig_phasediags} by the solid and dashed black lines, respectively.

To address the question of which path is taken in the phase diagram as the resonance
energy is varied, we consider the two limiting cases of an adiabatic sweep and a sudden change in $\epsilon_{\rm{res}}$. 

\subsection{Adiabatic sweep of $\epsilon_{\rm{res}}$}
If the resonance energy is varied on a timescale much longer than the time needed for the gas to collisionally relax to equilibrium, then the system will follow a path of constant entropy in the phase diagram. The total entropy of the gas is given by
\begin{equation}
S_{\rm{tot}}(T,\mu,\epsilon_{\rm{res}}) = 2S_a(T,\mu) + \tilde S_m(T,\mu,\epsilon_{\rm{res}}).
\end{equation}
Note that the condensed molecules do not contribute to the entropy in the system. Starting on the
right side of the phase diagrams at some initial temperature $T_i$ and a large positive value of $\epsilon_{\rm{res}}$, there are no molecules and the initial entropy is given by
\begin{equation}
S_{\rm{tot}}(T_i,\mu_i) = k_{\rm{B}}N \Big [ 4 \frac{\mathcal F_4(e^{{\mu_i}/{k_{\rm{B}}T_i}})}{\mathcal F_3(e^{{\mu_i}/{k_{\rm{B}}T_i}})} - \frac{\mu_i}{k_{\rm{B}}T_i}
\Big ].
\label{initS}
\end{equation}
The resonance energy $\epsilon_{\rm{res}}$ is slowly lowered to a new value $\epsilon_{{\rm{res}},f}$. The new equilibrium solution is obtained by solving for $T_f$ and $\mu_f$ that satisfy the constraints of the total population and entropy being conserved. Above $T_c$ this
corresponds to solving for $T_f$ and $\mu_f$ in the equations
\begin{eqnarray}
N_{\rm{tot}}(T_f,\mu_f, \epsilon_{{\rm{res}},f}) &=& N , \\
S_{\rm{tot}}(T_f,\mu_f,\epsilon_{{\rm{res}},f})&=&S_{\rm{tot}}(T_i,\mu_i).
\label{constS}
\end{eqnarray}
Below $T_c$, $\mu_f=\epsilon_{{\rm{res}},f}/2$, so only Eq.~\ref{constS} must be solved.

\begin{figure}  %[htbp]
\begin{center} \includegraphics[scale=0.3]{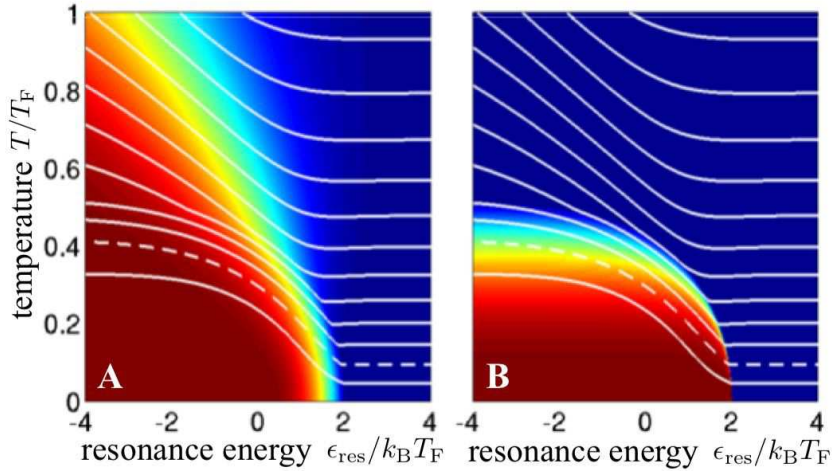}
\caption{Contours of constant entropy. Each solid white line defines the path followed by an adiabatic sweep of the resonance energy $\epsilon_{\rm{res}}$. For clarity, we plot the contours on top of the phase diagrams with the same color scale as in Figure 3. The dashed line corresponds to an initial temperature of $T_i/T_{\rm{F}} = 0.1$.}
\label{Fig_entropycontours}
\end{center}
\end{figure}
In Figure~\ref{Fig_entropycontours} we plot the contours of constant entropy (white lines) overlaying the phase diagrams. Each line defines the path followed by an adiabatic sweep of the resonance energy $\epsilon_{\rm{res}}$. Starting on the right side at a large value of $\epsilon_{\rm{res}}$, the temperature increases as $\epsilon_{\rm{res}}$ is lowered adiabatically, but eventually reaches a constant value when all of the atom pairs have been converted into molecules. We can understand the temperature increase intuitively: as pairs of atoms are converted into molecules, the system loses degrees of freedom, so the molecules must heat up to conserve entropy. Because an adiabatic process is reversible, it is possible to cool the gas by starting on the left side with all molecules and increasing $\epsilon_{\rm{res}}$. This useful idea was first pointed out by Carr {\it{et al.}} as a means to achieve a low enough temperature to get into the BCS region~\cite{Carr04a}. We emphasize that all of the atoms can be transfered into molecules if the sweep is done adiabatically.

\begin{figure}  %[htbp]
\begin{center} \includegraphics[scale=0.3]{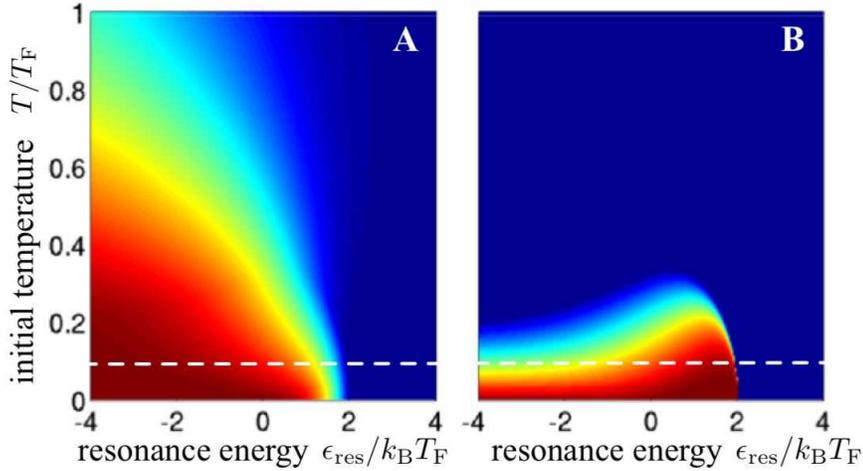}
\caption{Adiabatic phase diagrams. The molecule
fraction $\eta_m(\epsilon_{\rm{res}},S)$ is plotted in panel A and the condensate fraction
$\eta_{mc}(\epsilon_{\rm{res}},S)$ is shown in panel B. We label the $y$-axis, which corresponds to the entropy, by the initial temperature starting out with all atoms. The dashed white line corresponds to the dashed line of constant entropy in Figure~\ref{Fig_entropycontours}. }
\label{Fig_distorted}
\end{center}
\end{figure}
An alternative way to view the same information in Figure~\ref{Fig_entropycontours} is to plot
$\eta_m$ and $\eta_{mc}$ versus $\epsilon_{\rm{res}}$ and the {\it{entropy}} $S$, which we show in Figure~\ref{Fig_distorted}. We label the $y$-axes by the initial temperature $T_i$ starting with all atoms; the corresponding value of the initial entropy can be obtained from Eq.~(\ref{initS}) (where $\mu_i$ is chosen to give a total population of $N$). The resulting {\it{adiabatic}} phase diagrams look qualitatively different from the phase diagrams shown in Figure~\ref{Fig_phasediags}. In recent experiments, an adiabatic phase diagram was obtained by measuring the condensed molecule fraction after the magnetic field was adiabatically lowered starting with all atoms at some initial temperature $T_i$~\cite{Regal04a,Zwierlein04a}. It would be interesting to repeat these experiments and measure the {\it{final}} temperature of the gas after lowering the magnetic field, and thus obtain a phase diagram like the one shown in Figure~\ref{Fig_phasediags}.

\begin{figure}  %[htbp]
\begin{center} \includegraphics[scale=0.45]{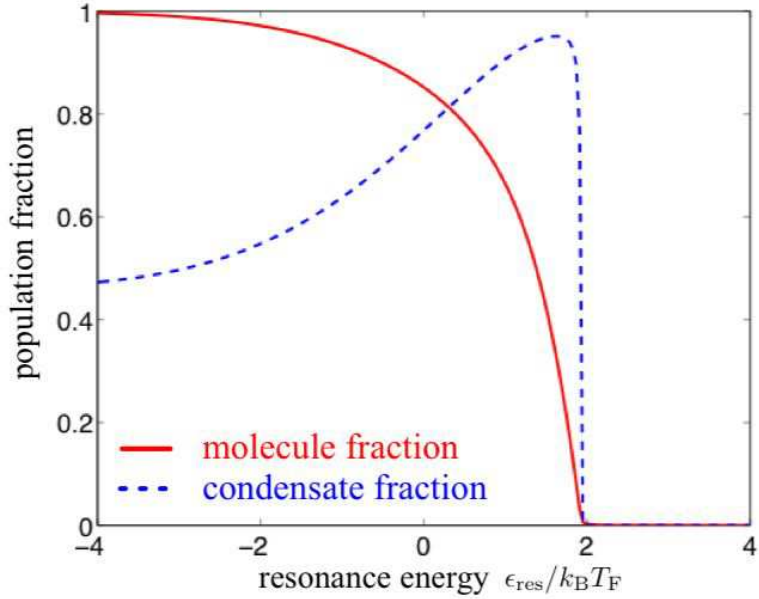}
\caption{Molecule fraction $\eta_m(T,\epsilon_{\rm{res}})$ (solid red line) and condensate fraction $\eta_{mc}(T,\epsilon_{\rm{res}})$ (blue dashed line) for an adiabatic ramp. These curves correspond to the dashed lines in Figures~\ref{Fig_entropycontours} and~\ref{Fig_distorted}, where the initial temperature is $T/T_{\rm{F}} = 0.1$ starting with all atoms.}
\label{Fig_entropypops}
\end{center}
\end{figure}

The dashed line in Figure~\ref{Fig_distorted} corresponds to the dashed contour shown in Figure~\ref{Fig_entropycontours} for an initial temperature of $T_i/T_{\rm{F}}=0.1$. In Figure~\ref{Fig_entropypops} we plot the molecule fraction and condensate fraction along this constant entropy contour. As $\epsilon_{\rm{res}}$ is lowered, the condensate fraction rises sharply and then decreases to some final value $\eta_{mc,f}$ after all the atoms have been converted to molecules. This same qualitative behavior was observed in recent experiments~\cite{Regal04a,Zwierlein04a}.

We can obtain analytic expressions in the high and low temperature limits for the final temperature and condensate fraction after all the atoms are converted to molecules. The initial entropy is $S_{\rm{tot}}(T_i)=2S_a(T_i,\mu_i)$, where $S_a$ is given in Eq.~(\ref{Sa}).
Using the Sommerfeld expansion in the low temperature limit, we find
\begin{equation}
S_{\rm{tot}}(T_i) \approx \pi^2 k_{\rm{B}}N \frac{T_i}{T_{\rm{F}}}.
\end{equation}
The final entropy is $S_{\rm{tot}}(T_f)=\tilde S_m(T_f,\epsilon_{\rm{res}}/2,\epsilon_{\rm{res}})$. Using expression Eq.~(\ref{Sm}) and taking $\mu_f = \epsilon_{\rm{res}}/2$, we find
\begin{equation}
S_{\rm{tot}}(T_f) = 4 \zeta(4) k_{\rm{B}} \Big (\frac{k_{\rm{B}}T_f}{\hbar \bar \omega} \Big )^3.
\end{equation}
Solving for $T_f$ in the equation $S_{\rm{tot}}(T_f)=S_{\rm{tot}}(T_i)$ and using $N = (k_{\rm{B}}T_{\rm{F}}/\hbar\bar\omega)^3/3$, the final temperature in the low temperature limit $T/T_{\rm{F}}\ll 1$ is
\begin{equation}
\frac{T_f}{T_{\rm{F}}}=\Big (\frac{\pi^2}{12 \zeta(4)} \Big )^{1/3} \Big (\frac{T_i}{T_{\rm{F}}} \Big )^{1/3}
\label{lowTf}
\end{equation}
The result $T_f \propto T_i^{1/3}$ agrees with the results of Carr {\it{et al.}}~\cite{Carr04a}. Their model treats mean field interactions in the condensate, so the constant of proportionality is different in the two models.

In the high temperature limit $T/T_{\rm{F}}\gg 1$, $\mathcal{F}_n(z)\approx z$ and $\mathcal{G}_n(z)\approx z$. Using this in the expressions for the entropy Eq.~(\ref{Sa}) and Eq.~(\ref{Sm}), and using Eq.~(\ref{Na}) and Eq.~(\ref{Nm}) to eliminate the chemical potentials, we obtain the initial and final entropies in the high temperature limit
\begin{eqnarray}
S_{\rm{tot}}(T_i) &=& k_{\rm{B}}N \Big \{ 4 - \ln \Big [ \Big (\frac{\hbar\bar\omega}{k_{\rm{B}}T_i}
\Big ) \frac{N}{2} \Big ] \Big \} , \\
S_{\rm{tot}}(T_f) &=& \frac{1}{2}k_{\rm{B}}N \Big \{ 4 - \ln \Big [ \Big (\frac{\hbar\bar\omega}{k_{\rm{B}}T_f}
\Big ) \frac{N}{2} \Big ] \Big \} .
\end{eqnarray}
Solving $S_{\rm{tot}}(T_f)=S_{\rm{tot}}(T_i)$, the final temperature in the high temperature limit $T/T_{\rm{F}}\gg 1$ is
\begin{equation}
\frac{T_f}{T_{\rm{F}}}=6^{1/3}e^{4/3}\Big(\frac{T_i}{T_{\rm{F}}} \Big )^2.
\label{highTf}
\end{equation}

\begin{figure}  %[htbp]
\begin{center} \includegraphics[scale=0.37]{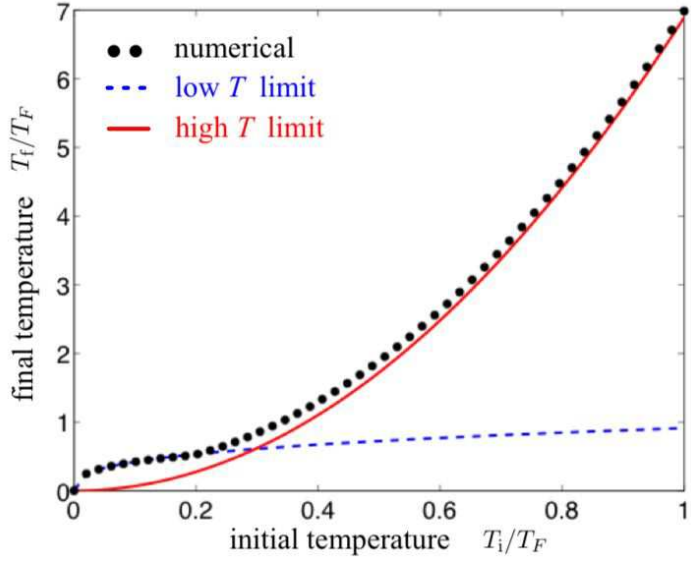}
\caption{Final temperature versus the initial temperature after all the atoms have been converted to molecules.}
\label{Fig_Tfinal}
\end{center}
\end{figure}

In Figure~\ref{Fig_Tfinal} we plot the final temperature as a function of the initial temperature. The black dotted line was obtained from a numerical solution. The solid red line corresponds to the high temperature limit Eq.~(\ref{highTf}) and the blue dashed line to the low temperature limit Eq.~(\ref{lowTf}). Both approximations agree well with the numerical solution in their appropriate regions of validity.

\begin{figure}  %[htbp]
\begin{center} \includegraphics[scale=0.375]{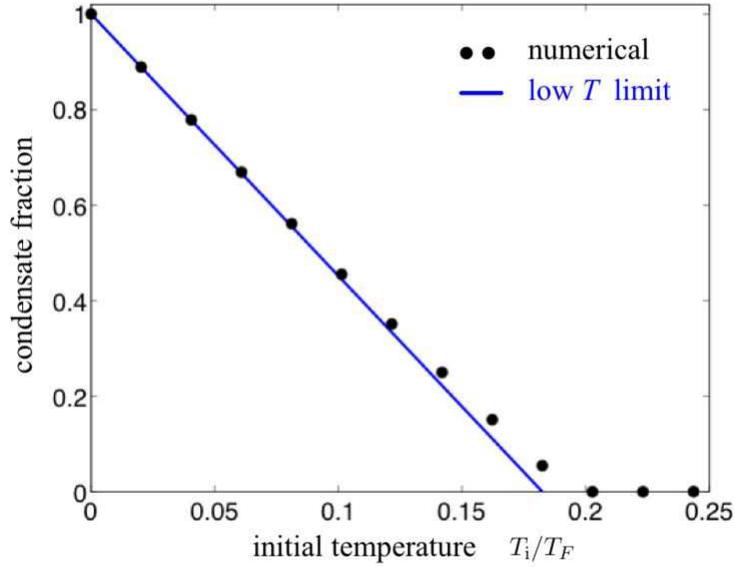}
\caption{Condensate fraction $\eta_{mc,f}$ after all the atoms have been converted to molecules versus the initial temperature.}
\label{Fig_Nfrac}
\end{center}
\end{figure}

We can calculate the condensate fraction at the end of the sweep in the limit of $|\epsilon_{{\rm{res}},f}|/k_{\rm{B}}T_{\rm{F}}\gg 1$ as a function of the intial temperature using the low temperature result in Eq.~(\ref{lowTf}). In terms of the final temperature, the condensate fraction is given by $\eta_{mc,f}=1 - (T_f/T_c)^3$, where $T_c=[6 \zeta(3)]^{-1/3}T_{\rm{F}}$. Substituting $T_f$ from
Eq.~(\ref{lowTf}), the condensate fraction in this limit varies linearly with the initial temperature as
\begin{equation}
\eta_{mc,f} = 1 - \frac{\pi^2 \zeta(3)}{2 \zeta(4)}\frac{T_i}{T_{\rm{F}}}.
\end{equation}
The critical {\it{initial}} temperature predicted by this equation, above which no condensed molecules exist when all the atoms have been converted to molecules, is $T_{i,c}/T_{\rm{F}} \approx 0.183$. In Figure~\ref{Fig_Nfrac} we plot the condensate fraction after an adiabatic sweep versus the initial temperature. We compare the low $T$ approximation to the full numerical solution. The deviation near $T_{i,c}$ arises from using only the first two terms in the Sommerfeld expansion in the low $T$ approximation.

Our simple model calculation helps to clarify the interpretation of experimental data concerning the critical temperature for superfluidity. In the BEC limit of negative detuning, the critical temperature observed in recent experiments is approximately $0.2  T_{\rm{F}}$~\cite{Regal04a,Zwierlein04a}. At first sight, this seems in contradiction with the prediction of $T_c/T_{\rm{F}} = 0.518$ for a trapped Bose gas~\cite{Ohashi2003a} and seems more consistent with the {\it{homogeneous}} gas result~\cite{Falco04a,Diener04a} of $T_c/T_{\rm{F}} = 0.218$. However, when one realizes that the experimental data is plotted versus the {\it{initial}} temperature, the observed value of $T_c/T_{\rm{F}}\approx0.2$ is consistent with the expected result for the critical initial temperature $T_{i,c}$ for a trapped ideal gas with an adiabatic sweep, as shown in Figure~\ref{Fig_Nfrac} and in the left side of Figure~\ref{Fig_distorted}B.

\subsection{Sudden change of $\epsilon_{\rm{res}}$}

In the previous section we considered the situation where $\epsilon_{\rm{res}}$ is varied on a timescale much longer than the collisional relaxation time for the gas to equilibrate. In this adiabatic limit, the gas dynamically follows along the constant entropy contours shown in Figure~\ref{Fig_entropycontours} and the process is reversible. We now consider the opposite limit of a sudden change in $\epsilon_{\rm{res}}$. In this case, it is the {\it{energy}} that is conserved rather than the entropy, and the process is {\it{not}} reversible. In the two sections below, we consider two different initial states, starting out with either a gas of only atoms (the right side of the phase diagram) or with a gas of only molecules (the left side of the phase diagram).

\subsubsection{Starting with all atoms}

Suppose the gas of fermionic atoms is cooled down to some temperature $T_i$ and that $\epsilon_{{\rm{res}},i}/k_{\rm{B}}T_{\rm{F}}\gg 1$, so that there are no molecules initially, ($\mu_i$ must be chosen to satisfy $N_a(T_i,\mu_i)=N/2$). The resonance energy is then suddenly lowered to a new value $\epsilon_{{\rm{res}},f}$ over a time much shorter than all collisional relaxation timescales. The initial total energy of the gas immediately following this jump in $\epsilon_{\rm{res}}$ is 
\begin{equation}
E_{\rm{tot}}(T_i,\mu_i)=2E_a(T_i,\mu_i),
\end{equation}
where $E_a(T,\mu)$ is defined in Eq.~(\ref{Ea}). The system subsequently relaxes from this initial state to a new equilibrium state with a different temperature $T_f$ and chemical potential $\mu_f$, during which time pairs of atoms may form molecules. The final energy is
\begin{eqnarray}
E_{\rm{tot}}(T_f,\mu_f,\epsilon_{{\rm{res}},f}) &=& 2E_a(T_f,\mu_f) + \tilde E_m(T_f,\mu_f,\epsilon_{{\rm{res}},f}) \nonumber \\ 
&+& \epsilon_{{\rm{res}},f}N_{mc}(T_f,\epsilon_{{\rm{res}},f}).
\label{Etotf}
\end{eqnarray}
Because the Hamiltonian for the system is conservative and does not depend on time after the sudden jump in $\epsilon_{\rm{res}}$, the final energy must be equal to the initial energy
\begin{equation}
E_{\rm{tot}}(T_f,\mu_f,\epsilon_{{\rm{res}},f})=E_{\rm{tot}}(T_i,\mu_i). 
\label{Econs}
\end{equation}
The values of $T_f$ and $\mu_f$ are chosen to satisfy Eq.~(\ref{Econs}) and number conservation.

\begin{figure} 
\begin{center} \includegraphics[scale=0.3]{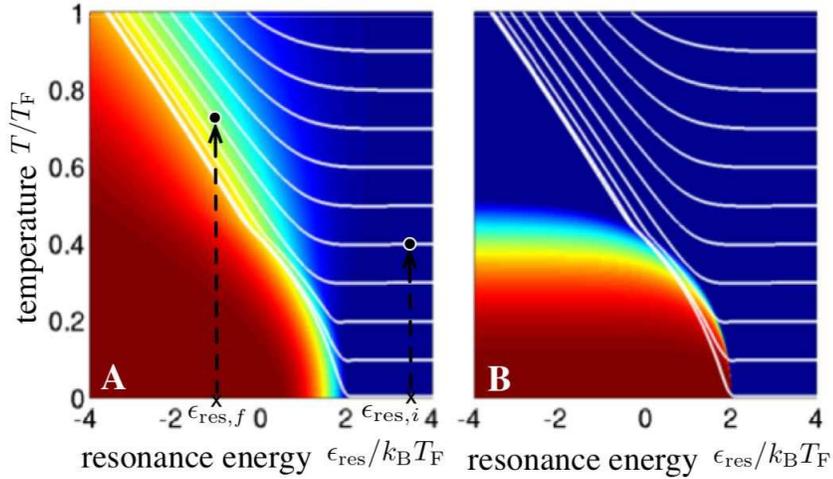}
\caption{Lines of constant energy starting with all atoms initially. The initial temperature for the lines start at $T_i=0$ and increase in increments of $0.1T_{\rm{F}}$. As in Figure~\ref{Fig_entropycontours}, we plot the contours on top of the phase diagrams showing the molecule fraction in panel A and the fraction of condensed molecules in panel B. For clarity, we show hypothetical intial and final points in panel A.}
\label{Fig_energycontours}
\end{center}
\end{figure}

In Figure~\ref{Fig_energycontours} we plot the lines of constant energy starting with all atoms initially. The initial temperature of the lines (on the right side of the plots) increases in steps of $0.1T_{\rm{F}}$, starting with the lowest line at zero temperature. The region below the lowest line is a forbidden region of the phase diagram that can not be reached by suddenly lowering $\epsilon_{\rm{res}}$. In Figure~\ref{Fig_energycontours}A we show a hypothetical final point starting at an intial temperature of $T_i=0.4T_{\rm{F}}$. We emphasize that the system does not dynamically follow along the path between the initial and final points, in contrast to the adiabatic case. After the sudden decrease in $\epsilon_{\rm{res}}$, the system evolves out of equilibrium away from the initial state and then relaxes to the final equilibrium state at a new temperature $T_f$. 
\begin{figure}  %[htbp]
\begin{center} \includegraphics[scale=0.45]{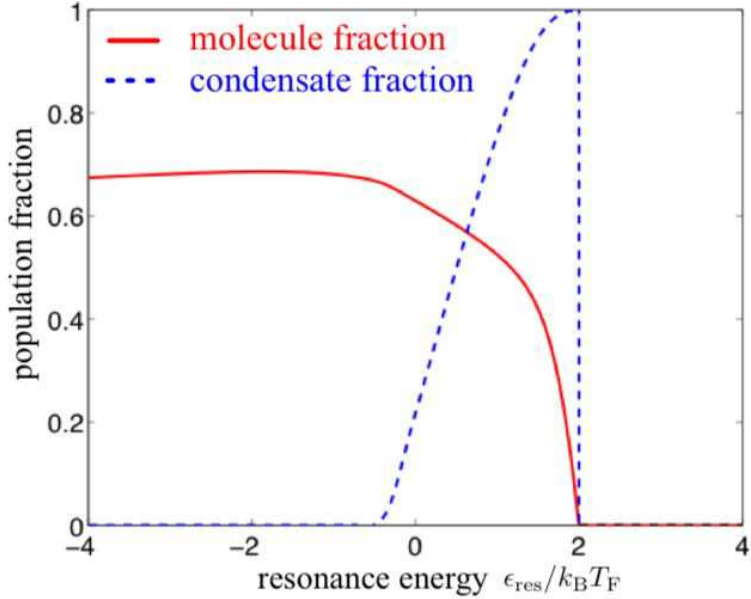}
\caption{Molecule fraction $\eta_m(T,\epsilon_{\rm{res}})$ (solid red line) and condensate fraction $\eta_{mc}(T,\epsilon_{\rm{res}})$ (blue dashed line) for a sudden ramp. These lines correspond to the bottom line shown in Figure~\ref{Fig_energycontours}, which starts at an initial temperature of $T_i = 0$.}
\label{Fig_energypops}
\end{center}
\end{figure}

From Figure~\ref{Fig_energycontours}A we can see that it is impossible to transfer all of the atoms to molecules by suddenly decreasing $\epsilon_{\rm{res}}$. This property was also noted by Chin and Grimm in Ref.~\cite{Chin03b} for a classical gas.  As  $\epsilon_{\rm{res}}$ decreases, the molecule fraction $\eta_m$ and condensate fraction $\eta_{mc}$ each reach a maximum and then decrease. These maximum values $\max \{\eta_m \}$ and $\max \{\eta_{mc} \}$ increase as the initial temperature $T_i$ is lowered. In Figure~\ref{Fig_energypops} we plot $\eta_m$ (solid red line) and $\eta_{mc}$ (blue dashed line) versus $\epsilon_{\rm{res}}$ for an initial temperature of $T_i=0$. At a value of $\epsilon_{\rm{res}}=2k_{\rm{B}}T_{\rm{F}}$, all of the molecules are condensed $\max \{\eta_{mc} \}=1$. As $\epsilon_{\rm{res}}$ is lowered, the temperature $T$ increases and the condensed fraction decreases sharply to zero. The molecule fraction increases to a maximum value of $\max \{\eta_m \}\approx 0.7$, then decreases slowly.

\subsubsection{Starting with all molecules}

We now consider the situation where we start out on the negative side of $\epsilon_{\rm{res}}$ with a gas of bosonic molecules at some initial temperature $T_i$ and that $|\epsilon_{{\rm{res}},i}|/k_{\rm{B}}T_{\rm{F}} \gg 1$, so that there are no atoms initially, ($\mu_i$ must be chosen to satisfy $\tilde N_m(T_i,\mu_i,\epsilon_{{\rm{res}},i})+N_{mc}(T_i,\mu_i,\epsilon_{{\rm{res}},i})=N/2$). The resonance energy is then suddenly raised to a new value $\epsilon_{{\rm{res}},f}$ over a time much shorter than all collisional relaxation timescales. The initial total energy of the gas immediately following this jump in $\epsilon_{\rm{res}}$ is 
\begin{equation}
E_{\rm{tot}}(T_i,\mu_i,\epsilon_{{\rm{res}},i})=3\tilde N_m(T_i,\mu_i,\epsilon_{{\rm{res}},i})   k_{\rm{B}} T_i \, \frac{\mathcal G_4(z_{m,i})}{\mathcal G_3(z_{m,i})} 
 + \frac{N}{2}\epsilon_{{\rm{res}},f} ,
\end{equation}
where $z_{m,i}=e^{2\mu_i+\epsilon_{{\rm{res}},i}}$ is the initial molecule fugacity. The first term describes the initial thermal energy of the gas. Note that it is the final value of $\epsilon_{{\rm{res}},f}$ appearing in the second term describing the internal molecular energy. The system subsequently relaxes to a new equilibrium state with a different temperature $T_f$ and chemical potential $\mu_f$, during which time some of the molecules may dissociate into pairs of atoms. The final energy is given by Eq.~(\ref{Etotf}).
The final values of $T_f$ and $\mu_f$ are chosen to satisfy $E_{\rm{tot}}(T_f,\mu_f,\epsilon_{{\rm{res}},f}) =E_{\rm{tot}}(T_i,\mu_i,\epsilon_{{\rm{res}},i})$ and number conservation.

\begin{figure}
\begin{center} \includegraphics[scale=0.385]{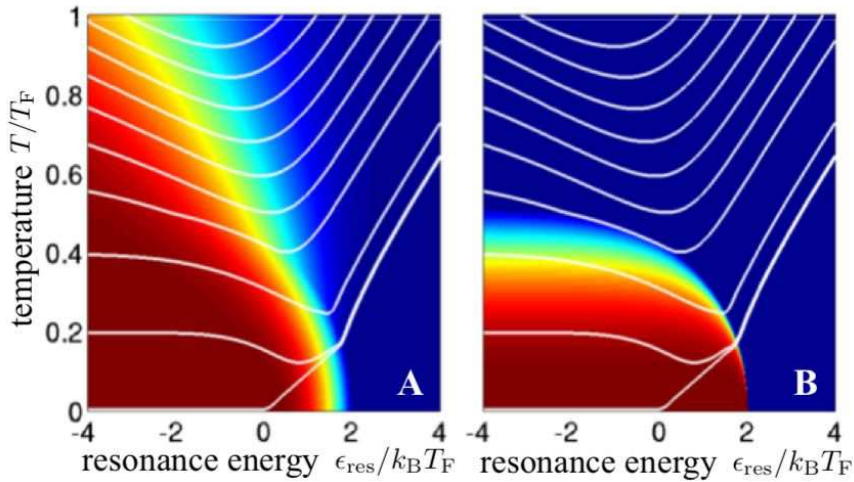}
\caption{Lines of constant energy starting with all molecules initially. The initial temperature for the lines start at $T_i=0$ and increase in increments of $0.2T_F$. The starting points of the upper lines lie out of the range shown in the graph.}
\label{Fig_energycontours2}
\end{center}
\end{figure}

In Figure~\ref{Fig_energycontours2} we plot the lines of constant energy starting with all molecules initially. The initial temperature of the lines increases in steps of $0.2T_{\rm{F}}$, starting with the lowest line at zero temperature. The region below the lowest line is a forbidden region of the phase diagram that can not be reached by suddenly raising $\epsilon_{\rm{res}}$. In the previous case of starting with all atoms, there is an energy barrier preventing a full transfer of all the atoms into molecules. Here we find that all of the molecules can dissociate as $\epsilon_{\rm{res}}$ is increased.

\section{Conclusions}

In this paper, we have studied the thermodynamics of molecule formation in an ideal gas mixture of fermionic atoms and bosonic molecules that are in chemical and thermal equilibrium, where the internal energy of the molecules can be adjusted relative to that of the atoms.  We considered both limiting cases of sweeping the resonance energy adiabatically or changing it suddenly. Our work is the natural extension into the quantum degenerate regime of the earlier work by Chin and Grimm~\cite{Chin03b} who studied a classical gas and has some overlap with the recent work of Carr {\it{et al.}}~\cite{Carr04a,Carr2004b}.

Our toy model serves as a useful guide for studying the problem of molecule formation in the adiabatic and sudden limits. In particular, we have given a prescription for calculating the adiabatic phase diagrams, shown in Figure~\ref{Fig_distorted}, in which the molecule fraction and fraction of condensed molecules are plotted versus the resonance position $\epsilon_{\rm{res}}$ and entropy $S$. These should be contrasted to the phase diagrams shown in Figure~\ref{Fig_phasediags}, where quantities are plotted versus temperature $T$ rather than entropy. In the experiments of Regal {\it{et al.}}~\cite{Regal04a} and Zwierlein {\it{et al.}}~\cite{Zwierlein04a}, the condensate fraction was measured as a function of the {\it{initial}} temperature and the magnetic field controling the resonance. The resulting graphs in Figure 4 of Ref.~\cite{Regal04a} and Figure 5 of Ref.~\cite{Zwierlein04a} are adiabatic phase diagrams, analogous to Figure~\ref{Fig_distorted}B; they are not the more intuitive phase diagram shown in Figure~\ref{Fig_phasediags}B~\cite{Falco04a,Diener04a}. This distinction is important to realize when comparing theory to experiment. 

As an example,  the theory of a trapped gas previously developed by Ohashi and Griffin~\cite{Ohashi2003a} predicted a BEC transition temperature of $T_c \approx 0.5 T_{\rm{F}}$. Subsequent treatments of a homogeneous gas by Falco and Stoof~\cite{Falco04a} and Diener and Ho~\cite{Diener04a} obtained a transition temperature of $T_c \approx 0.2 T_{\rm{F}}$. However, none of these theoretical values can be compared directly with any temperature that was actually measured in the experiments of Regal {\it{et al.}}~\cite{Regal04a} and Zwierlein {\it{et al}}~\cite{Zwierlein04a} because the only temperatures reported there are those of the Fermi gas at the start of the magnetic field sweep. The construction of the adiabatic phase diagram that is appropriate to compare our model with experimental data shows that an atomic gas with initial temperature of $T_i \approx 0.2 T_{\rm{F}}$ is adiabatically converted to a molecular gas with a temperature of $T_f \approx 0.5 T_{\rm{F}}$, which is equal to the critical temperature for a trapped ideal Bose gas. 

In order to make a detailed comparison with experiments, a more advanced many-body theory that properly treats the effect of the resonant interactions must be used in order to describe the crossover to BCS superfluidity in the positive detuning region~\cite{Ohashi02a,Milstein2002a,Falco04a,Diener04a,Stajic2004a,Ohashi2003a}. Furthermore, in experimental data the resonance position is typically expressed in terms of the external magnetic field $B$ rather than the energy of the resonant state $\epsilon_{\rm{res}}(B)$; this must also be taken into account when comparing theory to experiment.

\section{Acknowledgements}
We would like to thank Tetsuro Nikuni for useful discussions and Allan Griffin for helpful suggestions about the manuscript.

 \end{document}